\documentclass[11pt,a4paper]{article}
\usepackage{amsmath, amsfonts, amssymb}
\usepackage{a4wide}
\usepackage{parskip}
\usepackage{enumitem}
\usepackage{xcolor}

\usepackage{hyperref}
\usepackage{booktabs}
\usepackage{caption}
\usepackage{siunitx}
\usepackage{tabularx}
\usepackage{adjustbox}
\usepackage{multirow}
\usepackage{amsthm}
\usepackage{graphicx}
\usepackage{bm}
\usepackage{lscape}
\usepackage{comment}
\usepackage{float}
\usepackage{tabularray}
\UseTblrLibrary{amsmath}
\usepackage{nicematrix}
\usepackage[noadjust]{cite}

\newtheorem{theorem}{Theorem}[section]
\newtheorem{lemma}[theorem]{Lemma}
\theoremstyle{definition}
\newtheorem{definition}[theorem]{Definition}

\theoremstyle{remark}
\newtheorem{remark}[theorem]{Remark}

\numberwithin{equation}{section}

\newcommand{\Char}{\textnormal{char}}

\newcommand{\ord}{\textnormal{ord}}
\newcommand{\wt}{\textnormal{wt}}
\newcommand{\rank}{\textnormal{rank}}
\newcommand{\lpf}{\textnormal{lpf}}

\usepackage{tikz,xcolor,hyperref}

\definecolor{lime}{HTML}{A6CE39}
\DeclareRobustCommand{\orcidicon}{%
	\begin{tikzpicture}
		\draw[lime, fill=lime] (0,0) 
		circle [radius=0.16] 
		node[white] {{\fontfamily{qag}\selectfont \tiny ID}};
		\draw[white, fill=white] (-0.0625,0.095) 
		circle [radius=0.007];
	\end{tikzpicture}
	\hspace{-2mm}
}

\foreach \x in {A, ..., Z}{%
	\expandafter\xdef\csname orcid\x\endcsname{\noexpand\href{https://orcid.org/\csname orcidauthor\x\endcsname}{\noexpand\orcidicon}}
}

\begin{document}
	\date{}
	{\vspace{0.01in}
		\title{On the Euclidean duals of the cyclic codes generated via cyclotomic polynomials}
		\author{{\bf Anuj Kumar Bhagat\footnote{email: {\tt anujkumarbhagat632@gmail.com}}\orcidA{}\;and \bf Ritumoni Sarma\footnote{	email: {\tt ritumoni407@gmail.com}}\orcidB{}} \\ Department of Mathematics,\\ Indian Institute of Technology Delhi,\\Hauz Khas, New Delhi-110016, India.}
		\maketitle
		\begin{abstract} 
            For a natural number $n\ge2$ which is co-prime to $\Char(\mathbb{F}_q)$, let $\mathcal{C}_n$ and $\mathcal{C}_{n,1}$ denote the cyclic codes of length $n$ over $\mathbb{F}_q$ generated by the $n$-th cyclotomic polynomial $Q_n(x)$ and the polynomial $Q_n(x)Q_1(x)$, respectively. In \cite{BHAGAT2025}, the minimum distances of the codes $\mathcal{C}_n$ and $\mathcal{C}_{n,1}$ were determined, and a conjecture regarding the minimum distances of their Euclidean duals was proposed. In this article, we completely describe the structure of these dual codes and as a consequence, we find their minimum distances explicitly as functions of $n$. In fact, we resolve the conjecture in \cite{BHAGAT2025} by proving that the minimum distance of the Euclidean dual of each of $\mathcal{C}_n$ and $\mathcal{C}_{n,1}$ is equal to $2^{\omega(n)}$.
			\medskip
			
			
			\medskip
			
			
		\end{abstract}
\section{Introduction}\label{Section 1}
Linear codes play a central role in algebraic coding theory owing to their rich algebraic structure, which facilitates efficient computation. Cyclic codes, a prominent subclass introduced by Prange \cite{prange1957cyclic}, are invariant under cyclic shifts, endowing them with additional algebraic structure and enabling efficient encoding and decoding procedures. Consequently, cyclic codes find widespread applications in consumer electronics, digital storage, and data transmission systems. From an algebraic perspective, a cyclic code of length $n$ over $\mathbb{F}_q$ can be identified with an ideal of the quotient ring $\mathbb{F}_q[x]/\langle x^n - 1 \rangle$, generated by a divisor of $x^n - 1$. In particular, if $d \mid n$, then the $d$-th cyclotomic polynomial $Q_d(x)$ divides $x^n - 1$ in $\mathbb{F}_q[x]$, which naturally motivates the study of cyclic codes generated by cyclotomic polynomials. We denote by $\mathcal{C}_n$ the cyclic code generated by the $n$-th cyclotomic polynomial over $\mathbb{F}_q$, and by $\mathcal{C}_{n,1}$ the code generated by $Q_n(x)Q_1(x)$ over $\mathbb{F}_q$.

Determining the minimum distance of a code is fundamental for assessing its error-correcting capability. However, computing the exact minimum distance of a linear code is, in general, NP-hard \cite{Dumer_distance_hard}, even for structured families such as cyclic codes. Consequently, estimating the minimum distance of cyclic codes remains a classical and important problem in coding theory. Several lower bounds have been developed for this purpose, including the Bose--Chaudhuri--Hocquenghem (BCH), Hartmann--Tzeng (HT), Roos, and shift bounds (see \cite{bose1960class, van2003minimum, roos2003new, hartmann1972generalizations}). In \cite{BHAGAT2025}, the minimum distances of the cyclic codes $\mathcal{C}_n$ and $\mathcal{C}_{n,1}$ were determined as functions of their length $n$, and it was conjectured that the minimum distances of their Euclidean duals also depend only on $n$, specifically, $2^{\omega(n)}$. This conjecture was verified when $n$ has at most two distinct prime factors. In the present work, we resolve this conjecture in full generality.

A linear complementary dual (LCD) code is a linear code whose intersection with its dual is trivial. Such codes have important applications in data storage and cryptography. In \cite{Carlet}, Carlet and Guilley demonstrated that LCD codes provide effective countermeasures against side-channel and fault injection attacks. The code families $\mathcal{C}_n$ and $\mathcal{C}_{n,1}$ are both LCD.

Locally recoverable codes (LRCs) enable recovery from erasures by accessing only a small number of intact coordinates. To formalize this, the notion of locality was introduced in \cite{locality}, defined as the minimum number of symbols required to repair a coordinate. Although small locality is desirable, codes achieving the smallest locality do not necessarily attain optimal parameters with respect to the Singleton-like bound \cite{locality} or the Cadambe--Mazumdar bound~\cite{Cadambe}. LRCs have important applications in distributed storage systems, where they efficiently handle single-node failures, a common practical scenario. Several constructions of LRCs have been proposed in the literature. In particular, \cite{viaCycloPol} studied a general construction of $q$-ary cyclic LRCs using multiples of cyclotomic polynomials, while further constructions appear in \cite{LRCalgebraic, LRCJin, LRC_Optimal, LRCelliptic, TamoandBarg, Distance3and4, consta_dual, Consta_prime_power, Consta_LRC}. Cyclic LRC that is also LCD, referred to as cyclic LCD-LRC, were considered in \cite{Mahesh}. Such codes are particularly attractive for distributed storage systems, as they simultaneously provide efficient local repair and robustness against side-channel and fault injection attacks. In \cite{BHAGAT2025}, the locality of the codes $\mathcal{C}_n$ and $\mathcal{C}_{n,1}$ was determined, and a few of their subfamilies were shown to be optimal. Consequently, these families form cyclic LCD-LRCs.

The remainder of the article is organized as follows. Section~\ref{Preliminaries} reviews the necessary preliminaries. In Section~\ref{section 3}, we formulate the problem and motivate the study of cyclic codes generated by cyclotomic polynomials and their Euclidean duals. Section~\ref{section 4} characterizes, up to permutation equivalence, the Euclidean duals of $\mathcal{C}_n$ and $\mathcal{C}_{n,1}$, and determines their minimum distances for all $n$ for which these codes are defined. Section~\ref{conclusion} concludes the article.

\section{Preliminaries}\label{Preliminaries}
Throughout this article, let $ p $ be a prime and $ q = p^l $ for some $ l \in \mathbb{N} $. Denote by $ \mathbb{F}_q $ the finite field of order $ q $ so that $ \mathrm{char}(\mathbb{F}_q) = p $. It is well known that the multiplicative group $ \mathbb{F}_q^\times := \mathbb{F}_q \setminus \{0\} $ is cyclic; any generator $ \beta \in \mathbb{F}_q^\times $ is called a \emph{primitive element} of $ \mathbb{F}_q $.

Let $ n \in \mathbb{N} $ be such that $ p \nmid n $. For each $ i \in \mathbb{Z}_n $, the set
\[
C_i := \{\, i q^j \bmod n : j \in \mathbb{N} \cup \{0\} \,\}
\]
is called the \emph{cyclotomic coset} of $q$ in $\mathbb{Z}_n$ containing $i$.

For $ \beta \in \mathbb{F}_{q^m} $, the \emph{minimal polynomial} of $ \beta $ over $ \mathbb{F}_q $ is defined as the unique monic polynomial $ f(x) \in \mathbb{F}_q[x] $ of least degree such that $ f(\beta) = 0 $. This polynomial is irreducible over $ \mathbb{F}_q $.

The following theorem characterizes the minimal polynomial of an element in a finite field.
\begin{theorem}\cite{LING}
Let $\beta \in \mathbb{F}_{q^m}$ be a primitive element. If $C_i$ is the cyclotomic coset of $q$ in $\mathbb{Z}_{q^m-1}$ containing $i$, then $M^{(i)}(x):=\underset{j\in C_i}{\prod} (x-\beta^j)$ is the minimal polynomial of $\beta^i$ over $\mathbb{F}_q$.
\end{theorem}
\begin{theorem}\label{x^n-1}\cite{LING}
    Let $n\in\mathbb{N}$ be such that $\gcd(n,q)=1$ and let $t$ be the order of $q$ in $U(\mathbb{Z}_n).$ Suppose $\zeta$ is a primitive $n$-th root of unity in $\mathbb{F}_{q^t}.$ Then $x^n-1=\underset{s}{\prod}M^{(s)}(x),$ where $s$ runs through a set of representatives of the cyclotomic cosets of $q$ in $\mathbb{Z}_n$ and $M^{(s)}(x)$ is the minimal polynomial of $\zeta^s$ over $\mathbb{F}_q.$
\end{theorem}
A \textit{linear code} $ \mathcal{C} $ of length $ n $ and dimension $ k $ over $ \mathbb{F}_q $ is a $ k $-dimensional subspace of the vector space $ \mathbb{F}_q^n $. The elements of $ \mathcal{C} $ are called \textit{codewords}. The \textit{(Hamming) weight} of a codeword $ \bm{c} = (c_0, \dots, c_{n-1}) \in \mathcal{C} $ is defined by
\[
\mathrm{wt}(\bm{c}) := \#\{\, 0 \le i \le n-1 : c_i \ne 0 \,\}.
\]
The \textit{(Hamming) distance} of $ \mathcal{C} $ is given by
\[
d(\mathcal{C}) = \min \{\, \mathrm{wt}(\bm{c}) : \bm{c} \in \mathcal{C} \setminus \{\bm{0}\} \,\}.
\]
A linear code $ \mathcal{C} $ of length $ n $ and dimension $ k $ over $ \mathbb{F}_q $ is called an $ [n,k]_q $-code. If the minimum distance $ d(\mathcal{C}) $ is known, then $ \mathcal{C} $ is called an $ [n,k,d]_q $-code.

An $[n,k,d]_q$-code $ \mathcal{C} $ over $ \mathbb{F}_q $ is called \textit{cyclic} if for every $ \bm{c} = (c_0, c_1, \dots, c_{n-1}) \in \mathcal{C} $, the vector $ (c_{n-1}, c_0, \dots, c_{n-2}) $ also belongs to $ \mathcal{C} $.

A cyclic code $ \mathcal{C} $ can be identified with an ideal of the ring $ \mathbb{F}_q[x]/\langle x^n - 1 \rangle $ by associating each vector $ \bm{c} = (c_0, c_1, \dots, c_{n-1}) \in \mathbb{F}_q^n $ with the residue class
\[
\bm{c}(x)+\langle x^n - 1 \rangle \in \mathbb{F}_q[x]/\langle x^n - 1 \rangle,
\]
where $\bm{c}(x):= c_0 + c_1 x + \cdots + c_{n-1} x^{n-1}.$ For simplicity, we denote this residue class by $ \bm{c}(x) $.

Under this identification, and since $ \mathbb{F}_q[x]/\langle x^n - 1 \rangle $ is a principal ideal ring, the cyclic code $ \mathcal{C} $ is generated by a unique monic polynomial $ g(x) $ dividing $ x^n - 1 $. We write $ \mathcal{C} = \langle g(x) \rangle $ and call $ g(x) $ the \textit{generator polynomial} of $ \mathcal{C} $.
 \begin{theorem}\cite{LING}
     Let $\mathcal{C}$ be a cyclic code of length $n$ over $\mathbb{F}_q$ generated by $g(x).$ Then $\dim \mathcal{C}=n-\deg g(x).$
 \end{theorem}
Let $\gcd(n,q)=1.$ Suppose that $\mathcal{C}=\langle g(x) \rangle $ is a cyclic code of length $n$ over $\mathbb{F}_q$. By Theorem \ref{x^n-1},
$$
g(x)=\underset{s\in \mathcal{A}}{\prod}\, \,\underset{i\in C_s}{\prod}(x-\zeta^i),
$$
where $s$ runs through some subset $\mathcal{A}$ of
representatives of the cyclotomic cosets $C_s$ of $q$ in $\mathbb{Z}_n$. Let $T=\underset{s\in \mathcal{A}}{\bigcup}\,  C_s.$ The roots of unity $Z =\{\zeta^i: i\in T \}$ are called the \textit{zeros} of the
cyclic code $\mathcal{C}$ and $\{\zeta^i: i\notin T \}$ are called the \textit{nonzeros} of $\mathcal{C}.$ The set $T$ is called the \textit{defining set} of $\mathcal{C}.$ 

Moreover, the nonzeros of $ \mathcal{C} $ are precisely the roots of the polynomial $ h(x) = \frac{x^n - 1}{g(x)} $.
\begin{definition}
    The \textit{(Euclidean) dual} of an $[n,k]_q$-code $\mathcal{C}$ over $\mathbb{F}_q$ is the linear code defined by
    $$
        \mathcal{C}^{\perp}:=\{\bm{x}\in \mathbb{F}_q^n: \bm{x}\cdot \bm{c}=0, \forall\, \bm{c}\in \mathcal{C}\},
    $$
    where $\bm{x}\cdot \bm{c}$ is the Euclidean inner product on $\mathbb{F}_q^n.$
\end{definition}
\begin{definition}
    Given a polynomial $f(x)\in \mathbb{F}_q[x],$ the polynomial $f^*(x):=x^{\deg f(x)}f\left(\frac{1}{x}\right)$ is called the \textit{reciprocal polynomial} of $f.$ Moreover, if $f=f^*,$ then $f$ is called \textit{self-reciprocal.}
\end{definition}
\begin{remark}
     The dual of an $[n,k]_q$-cyclic code is an $[n,n-k]_q$-cyclic code. If $\mathcal{C}=\langle g(x)\rangle$ and $h(x):=(x^n-1)/g(x),$ then $\mathcal{C}^{\perp}=\langle g^{\perp}(x)\rangle,$ where $g^{\perp}(x)={h^*(x)}/{h(0)}.$
\end{remark}
\begin{definition}
    Let $\mathcal{C}_i$ be an $[n_i, k_i, d_i]_q$-code over $\mathbb{F}_q$, for $i=1, 2$. Then the \textit{direct sum} of the codes $\mathcal{C}_1$ and $\mathcal{C}_2$ is defined as follows:
    \begin{equation}
        \mathcal{C}_1\oplus \mathcal{C}_2:=\{ (\bm{c}_1, \bm{c}_2): \bm{c}_1\in\mathcal{C}_1, \bm{c}_2\in\mathcal{C}_2\}.
    \end{equation}
\end{definition}
\begin{remark}
    It is a well-known fact that $\mathcal{C}_1\oplus \mathcal{C}_2$ is an $[n_1+n_2, k_1+k_2, \min \{d_1, d_2\}]_q$-code over $\mathbb{F}_q$. Moreover, if $G_i$ is a generator matrix of $\mathcal{C}_i$, for $i=1,2$, then a generator matrix of $\mathcal{C}_1\oplus \mathcal{C}_2$ is given by
    \begin{equation}
        \begin{bmatrix}
            G_1 & O\\
            O & G_2\\
        \end{bmatrix}
    \end{equation}
\end{remark}
\begin{definition}
    For $n\in \mathbb{N},$ the code $\mathcal{R}_n:=\{(\underbrace{\lambda,\lambda,\dots,\lambda}_{n\ \text{times}}): \lambda\in\mathbb{F}_q\}$ is called a \textit{repetition code} of length $n$ over $\mathbb{F}_q$.
\end{definition}
\begin{remark}
    It is clear that $\mathcal{R}_n$ is an $[n, 1, n]_q$-linear code over $\mathbb{F}_q$. Moreover, its Euclidean dual has parameters $[n, n-1, 2]_q$.  
\end{remark}
\begin{lemma}\cite{FIELD}\label{Existenceofe}
Let $g(x)\in\mathbb{F}_q[x]$ be a polynomial of degree $m$ such that $x\nmid g(x)$. Then for some integer $e$ with $1\leq e\leq q^m-1$, $g(x)$ divides $x^e-1$.
\end{lemma}
Lemma \ref{Existenceofe} motivates the following definition.
\begin{definition}\cite{FIELD}
Let $g(x)\in \mathbb{F}_q[x]\setminus \{0\}$. If $x\nmid g(x)$, then the least natural number $e$ such that $g(x)$ divides $x^e-1$ is called the \emph{order} of $g$ and it is denoted by ord$(g)$ or ord$(g(x))$. If $g(x)=x^rf(x)$ for some $r\in\mathbb{N}$ and $f(x)\in\mathbb{F}_q[x]$ with $x\nmid f(x)$, then ord$(g)$ is defined to be ord$(f)$.
\end{definition}
\begin{lemma}\cite{FIELD}\label{Ord divides c iff condition}
    Let $c$ be a positive integer. Then $f(x)\in\mathbb{F}_q[x]$ with $f(0)\ne 0$ divides $x^c- 1$ if and only if $\ord(f)$ divides $c.$
\end{lemma}
Throughout this article, for any positive integer $n>1,$ $\omega(n)$ denotes the number of distinct prime factors of $n$ and $\lpf(n)$ denotes the least prime factor of $n.$

\section{Problem Statement}\label{section 3}
In this short section, we introduce the problem in brief.

By Lemma \ref{Ord divides c iff condition}, for any cyclic code $\mathcal{C} = \langle g(x) \rangle$ of length $n$ over $\mathbb{F}_q$, the order of $g(x)$ divides $n$, that is, $\ord(g) \mid n$.
\begin{lemma}\label{e<n implies d<=2}
    Let $\mathcal{C}=\langle g(x)\rangle$ be a cyclic code of length $n$ over $\mathbb{F}_q.$ If $\ord(g)<n,$ then $d(\mathcal{C})\le 2.$ 
\end{lemma}
\begin{proof}
    If $e:=\ord(g),$ then $g(x)|x^e-1.$ Since $e<n,$ $x^e-1\in \mathcal{C}.$ Consequently, $d(\mathcal{C})\le \wt(x^e-1)=2.$  
\end{proof}
Lemma \ref{e<n implies d<=2} shows that a necessary condition for obtaining cyclic codes with a minimum distance of at least $3$ is that the length of the cyclic code must be equal to the order of its generator polynomial. Henceforth, for any polynomial $g(x) \in \mathbb{F}_q[x]$ with $g(0) \ne 0$, we consider the cyclic code $\langle g(x) \rangle$ of length $\ord(g)$ over $\mathbb{F}_q.$

The roots of the polynomial $x^n-1\in \mathbb{F}_q[x]$ (in the splitting field of $x^n-1$) are called the \textit{$n$-th roots of unity} over $\mathbb{F}_q$. Suppose $n\in \mathbb{N}$ is such that $\Char(\mathbb{F}{_q})\nmid n.$ Then the $n$th roots of unity form a cyclic group under multiplication. A generator of this cyclic group is called a \textit{primitive $n$th root of unity} over $\mathbb{F}_q.$
\begin{definition}
     Suppose $n\in \mathbb{N}$ is such that $\Char(\mathbb{F}{_q})\nmid n$ and $\zeta$ is a primitive $n$-th root of unity. Then the \textit{$n$th cyclotomic polynomial} over $\mathbb{F}_q$ is the polynomial
     $$
        Q_n(x):=\underset{\underset{\gcd(s,n)=1}{1\le s\le n}}{\prod} (x-\zeta^s).
     $$
\end{definition}
\begin{remark}
    By definition, $Q_n(x)$ is monic and $\deg Q_n(x)=\varphi(n),$ where $\varphi(n)$ is the Euler's Totient function.
\end{remark}
\begin{theorem}\label{factor x^n-1}\cite{FIELD}
    Suppose $n\in \mathbb{N}$ is such that $\Char(\mathbb{F}{_q})\nmid n.$ Then $Q_n(x)\in \mathbb{F}_p[x]$ and $x^n-1=\underset{d\mid n}{\prod} Q_d(x)$, wher $p=\Char({\mathbb{F}_q})$.
\end{theorem}
It is clear from Theorem \ref{factor x^n-1} that $Q_n(x)$ is a factor of $x^n-1.$ The next lemma shows that $n$ is the least value of $e$ such that $Q_n(x)$ is a factor of $x^e-1,$ for $e\in\mathbb{N}$.
\begin{lemma}\label{order of Q_n}
    Suppose $n\in \mathbb{N}$ is such that $\Char(\mathbb{F}{_q})\nmid n.$ Then the cyclotomic polynomial $Q_n(x)\in \mathbb{F}_q[x]$ has order $n$.
\end{lemma}
It is also evident that the order of $Q_n(x)Q_1(x)$ equals $n$. Consequently, it is natural to consider the cyclic codes $\mathcal{C}_{n}$ and $\mathcal{C}_{n,1}$, both of length $n$, generated by $Q_n(x)$ and $Q_n(x)Q_1(x)$, respectively. In \cite{BHAGAT2025}, the authors analyzed these codes and established their minimum distances, as presented in the following theorems.
\begin{theorem}\label{C_n}\cite{BHAGAT2025}
    Let $n>1$ be a positive integer such that $\Char(\mathbb{F}_q) \nmid n.$ Then the cyclic code $\mathcal{C}_n$ generated by $Q_n(x)\in \mathbb{F}_q[x]$ is an $[n,n-\varphi(n), \lpf(n)]_q$-code over $\mathbb{F}_q$.
\end{theorem}
\begin{theorem}\label{C_{n,1}}\cite{BHAGAT2025}
    Let $n$ be a composite number such that $\Char(\mathbb{F}_q) \nmid n.$ Then the cyclic code $\mathcal{C}_{n,1}$ generated by $Q_n(x)Q_1(x)\in \mathbb{F}_q[x]$ is an $[n,n-\varphi(n)-1, 2\times \lpf(n)]_q$-code over $\mathbb{F}_q$.
\end{theorem}
\begin{remark}
    If $n$ is prime, then $\mathcal{C}_{n,1}$ is the zero code since $Q_n(x)Q_1(x)=x^n-1.$ Therefore, throughout this article, we restrict our attention to composite values of $n$ while studying the code $\mathcal{C}_{n,1}$ and its Euclidean dual.
\end{remark}
Theorems \ref{C_n} and \ref{C_{n,1}} show that the minimum distances of the codes $\mathcal{C}_n$ and $\mathcal{C}_{n,1}$ are functions of $n$. It is therefore natural to inquire about the minimum distances of their Euclidean duals. In \cite{BHAGAT2025}, we conjectured that the minimum distances of their Euclidean duals are also functions of $n$, specifically $2^{\omega(n)}$, and proved the conjecture for the case when $n$ is the product of two distinct prime powers. We establish these conjectures in the next section.
\section{Euclidean Dual of the codes $\mathcal{C}_n$ and $\mathcal{C}_{n,1}$}\label{section 4}
In this section, we determine, up to permutation equivalence, the structure of the Euclidean dual of the codes $\mathcal{C}_n$ and $\mathcal{C}_{n,1}$, and as a result conclude that their minimum distance is $2^{\omega(n)}$. We begin by recalling the notion of the direct product of codes, as introduced in \cite[Chapter 18]{macwilliams1977theory}.
\begin{definition}\cite{macwilliams1977theory}
    Given two linear codes $\mathcal{C}_1$ and $\mathcal{C}_2$ over $\mathbb{F}_q$ with parameters $[n_1, k_1, d_1]$ and $[n_2, k_2, d_2]$ respectively, the \textit{direct product} of $\mathcal{C}_1$ and $\mathcal{C}_2,$ denoted by $\mathcal{C}_1\otimes \mathcal{C}_2$ is a code whose codewords are $n_1\times n_2$ arrays such that its columns are codewords of $\mathcal{C}_1$ and its rows are codewords of $\mathcal{C}_2.$ The code $\mathcal{C}_1\otimes \mathcal{C}_2$ is an $[n_1n_2, k_1k_2, d_1d_2]$-linear code over $\mathbb{F}_q.$
\end{definition}
A typical element of $\mathcal{C}_1\otimes \mathcal{C}_2$ is given by
\begin{equation}\label{Eqn:: typical element of C_1 otimes C_2}
    \begin{bmatrix}
        c_{0, 0} & c_{0, 1} & \cdots & c_{0, n_2-1} \\
        c_{1, 0} & c_{1, 1} & \cdots & c_{1, n_2-1} \\
        \vdots & \vdots & \cdots& \vdots \\
        c_{n_1-1, 0} & c_{n_1-1, 1} & \cdots & c_{n_1-1, n_2-1} \\
    \end{bmatrix},
\end{equation}
where $(c_{0, j}, c_{1, j}, \dots, c_{n_1-1, j})\in \mathcal{C}_1,$ for all $j\in \{0, 1, 2, \dots, n_2-1\}$ and  $(c_{i, 0}, c_{i, 1},  \dots, c_{i, n_2-1})\in\mathcal{C}_2,$ for all $i\in \{0, 1, 2, \dots, n_1-1\}.$
\begin{remark}
    \begin{enumerate}
        \item[(1)] It is easy to prove that if $G_1$ and $G_2$ are the generator matrices of the codes $\mathcal{C}_1$ and $\mathcal{C}_2,$ then $G_1\otimes G_2$ is a generator matrix of $\mathcal{C}_1\otimes\mathcal{C}_2,$ where $G_1\otimes G_2$ denotes the Kronecker product of matrices.
        \item[(2)] For two matrices $A$ and $B$ over a field, the code generated by the matrix $A\otimes B$ and the code generated by the matrix $B\otimes A$ are permutation equivalent.
    \end{enumerate}
\end{remark}
Suppose that the codes $\mathcal{C}_1$ and $\mathcal{C}_2$ are both cyclic, then $\mathcal{C}_1\otimes \mathcal{C}_2$ is closed under cyclic shift of all rows simultaneously and cyclic shift of all columns simultaneously. Algebraically, if $\mathcal{C}_1$ is an ideal of $\mathbb{F}_q[x]/\langle x^{n_1}-1\rangle$ and $\mathcal{C}_2$ is an ideal of $\mathbb{F}_q[y]/\langle y^{n_2}-1\rangle$, and if we identify a typical element in \eqref{Eqn:: typical element of C_1 otimes C_2} of $\mathcal{C}_1\otimes \mathcal{C}_2$ by
\begin{equation}
    f(x,y)=\sum_{i=0}^{n_1-1} \sum_{j=0}^{n_2-1} c_{i, j} x^i y^j \in \mathbb{F}_q[x,y]/\langle x^{n_1}-1, y^{n_2}-1\rangle, 
\end{equation}
then $xf(x,y)$ and $yf(x,y)$ represent the cyclic shift of all rows simultaneously and of all columns simultaneously, respectively. Set $\mathcal{R}_{n_1, n_2}:=\mathbb{F}_q[x,y]/\langle x^{n_1}-1, y^{n_2}-1\rangle$ and assume that $\gcd (n_1, n_2)=1.$ By the Chinese Remainder Theorem, for each $(i,j)\in \mathbb{Z}_{n_1}\times \mathbb{Z}_{n_2},$ there is a unique $\psi(i, j)\in \mathbb{Z}_{n_1 n_2}$ such that 
\begin{align}\label{CRT}
    \begin{split}
        \psi(i, j)&\equiv i \bmod n_1\\
        \psi(i, j)&\equiv j \bmod n_2.
    \end{split}
\end{align} 
Therefore, we have a ring isomorphism,
\begin{align}
    \begin{split}
        \Psi: \mathcal{R}_{n_1, n_2}&\to\mathbb{F}_q[z]/\langle z^{n_1n_2}-1 \rangle\\
        \Psi(x^iy^j)&=z^{\psi(i,j)}.
    \end{split}
\end{align}
Since $\psi(1, 1)=1,$ we have $\Psi(xy)=z$. Thus, for any codeword $f(x,y)\in\mathcal{C}_1\otimes\mathcal{C}_2$, we have $g(z):=\Psi(f(x,y))\in \Psi(\mathcal{C}_1\otimes\mathcal{C}_2)$. Since $f(x,y)\in \mathcal{C}_1\otimes \mathcal{C}_2$ implies $xyf(x,y)\in\mathcal{C}_1\otimes\mathcal{C}_2$, $g(z)\in \Psi(\mathcal{C}_1\otimes \mathcal{C}_2)$ implies $\Psi(xyf(x,y))= zg(z)\in \Psi(\mathcal{C}_1\otimes \mathcal{C}_2)$. This shows that $\Psi(\mathcal{C}_1\otimes \mathcal{C}_2)$ is cyclic. Moreover, $\mathcal{C}_1\otimes \mathcal{C}_2$ is permutation equivalent to $\Psi(\mathcal{C}_1\otimes \mathcal{C}_2).$ This gives the following result.
\begin{theorem} (\cite{macwilliams1977theory}, Chapter $18,$ Theorem $1$)
    If $\mathcal{C}_1$ and $\mathcal{C}_2$ are both cyclic codes of lengths $n_1$ and $n_2,$ respectively over $\mathbb{F}_q$ with $\gcd (n_1, n_2)=1,$ then $\mathcal{C}_1\otimes \mathcal{C}_2$ is permutation equivalent to a cyclic code. 
\end{theorem}
\begin{lemma}(\cite{macwilliams1977theory}, Chapter $18,$ Problem $5$)\label{Lem:: set of nonzeros}
    Let $\mathcal{C}_1$ and $\mathcal{C}_2$ be $[n_1, k_1]$ and $[n_2, k_2]$ cyclic codes, respectively, over $\mathbb{F}_q$ such that $\gcd (n_1, n_2)=1.$ If $\{\alpha_1, \alpha_2, \dots, \alpha_{k_1}\}$ and $\{\beta_1, \beta_2, \dots, \beta_{k_2}\}$ are respectively the set of nonzeros of $\mathcal{C}_1$ and $\mathcal{C}_2,$ then $\{\alpha_i \beta_j: 1\le i\le k_1, \; 1\le j\le k_2\}$ is the set of nonzeros of $\Psi (\mathcal{C}_1\otimes \mathcal{C}_2).$
\end{lemma}
\begin{proof}
    Let $\alpha^{n_1}=1$ and $\beta^{n_2}=1.$ For $f(x,y)=\sum_{i=0}^{n_1-1} \sum_{j=0}^{n_2-1} c_{i, j} x^i y^j \in\mathcal{C}_1\otimes\mathcal{C}_2$,
    \begin{align*}
        \Psi(f(x,y))(\alpha\beta)&=\sum_{i=0}^{n_1-1} \sum_{j=0}^{n_2-1} c_{i, j} (\alpha\beta)^{\psi(i,j)}\\
        &=\sum_{i=0}^{n_1-1} \sum_{j=0}^{n_2-1} c_{i, j} \alpha^{\psi(i,j)}\beta^{\psi(i,j)}\\
        &=\sum_{i=0}^{n_1-1} \sum_{j=0}^{n_2-1} c_{i, j} \alpha^i \beta^j\\
        &=f(\alpha, \beta),
    \end{align*}
    where the second last equality follows from \eqref{CRT}.
    
    Thus, $f(\alpha, \beta)=0$ if and only if $\Psi(f(x,y))(\alpha\beta)=0.$
\end{proof}
We now turn our attention to $\mathcal{C}_{n}^{\perp}$, the Euclidean dual of $\mathcal{C}_n := \langle Q_n(x) \rangle$. It was shown in \cite{BHAGAT2025} that $d(\mathcal{C}_{p_1^{a_1}}^{\perp}) = 2$, where $p_1$ is a prime co-prime to $\Char(\mathbb{F}_q)$ and $a_1 \in \mathbb{N}$. In the following theorem, we describe the structure of the code $\mathcal{C}_{p_1^{a_1}}^{\perp}$ which in turn establishes that $d(\mathcal{C}_{p_1^{a_1}}^{\perp}) = 2$.
\begin{theorem}\label{thm:: n is a single prime power}
    Let $p_1$ is a prime co-prime to $\Char(\mathbb{F}_q)$ and $a_1 \in \mathbb{N}$. Then the code $\mathcal{C}_{p_1^{a_1}}^{\perp}$ is permutation equivalent to the code $$\underbrace{\mathcal{C}_{p_1}^{\perp}\oplus\mathcal{C}_{p_1}^{\perp}\oplus\cdots \oplus \mathcal{C}_{p_1}^{\perp}}_{p_1^{a_1-1}\;\textnormal{copies}}.$$ In particular, $d(\mathcal{C}_{p_1^{a_1}}^{\perp})=2$.
\end{theorem}
\begin{proof}
    Since $Q_{p_1}(x)=1+x+x^2+\cdots+x^{p_1-1},$ $\mathcal{C}_{p_1}$ is a repetition code with parameters $[p_1, 1, p_1].$ Therefore, $\mathcal{C}_{p_1}^{\perp}$ is an $[p_1, p_1-1, 2]$-code. Now,
    \begin{align*}
        Q_{p_1^{a_1}}(x)&=Q_{p_1}\left(x^{{p_1}^{a_1-1}}\right)\\
        &=1+ x^{{p_1}^{a_1-1}}+ x^{2 {p_1}^{a_1-1}}+\cdots+x^{(p_1-1) {p_1}^{a_1-1}},
    \end{align*}
    which corresponds to the codeword $(1, \underbrace{0,0,\dots,0}_{p_1^{a_1-1}-1\ \text{times}}, 1,\underbrace{0,0,\dots,0}_{p_1^{a_1-1}-1\ \text{times}},\cdots, 1,\underbrace{0,0,\dots,0}_{p_1^{a_1-1}-1\ \text{times}})\in \mathcal{C}_{p_1^{a_1}}$ of weight $p$. Also, $\dim \mathcal{C}_{p_1^{a_1}}=p_1^{a_1}-\varphi(p_1^{a_1})=p_1^{a_1-1}.$ Thus, a generator matrix of $\mathcal{C}_{p_1^{a_1}}$ is given by
\begin{align*}
G &= \begin{bmatrix}
    Q_{p_1^{a_1}}(x)\\
    xQ_{p_1^{a_1}}(x)\\
    x^2 Q_{p_1^{a_1}}(x)\\
    \vdots \\
    x^{p_1^{a_1-1}-1}Q_{p_1^{a_1}}(x)
\end{bmatrix}\\
&=\begin{+bmatrix}
1 & 0 & 0 & \cdots & 0 & 1 & 0 & 0 & \cdots & 0 & \cdots & 1 & 0 & 0 & \cdots & 0 \\
0 & 1 & 0 & \cdots & 0 & 0 & 1 & 0 & \cdots & 0 & \cdots & 0 & 1 & 0 & \cdots & 0 \\
\vdots & \vdots & \vdots & \ddots & \vdots & \vdots & \vdots & \vdots & \ddots & \vdots & \cdots & \vdots & \vdots & \vdots & \ddots & \vdots \\
0 & 0 & 0 & \cdots & 1 & 0 & 0 & 0 & \cdots & 1 & \cdots & 0 & 0 & 0 & \cdots & 1
\end{+bmatrix}\\
&=\begin{bmatrix}
    I_{p_1^{a_1-1}} &\mid & I_{p_1^{a_1-1}} &\mid &\cdots & \mid & I_{p_1^{a_1-1}}
\end{bmatrix}_{p_1^{a_1-1}\times p_1^{a_1}}\\
&=\begin{bmatrix}
    1 & 1 & \cdots & 1
\end{bmatrix}\otimes I_{p_1^{a_1-1}},
\end{align*}
where $I_{p_1^{a_1-1}}$ denotes the identity matrix of order $p_1^{a_1-1}$.

Let $H:=H_{p_1}\otimes I_{p_1^{a_1-1}},$ where $H_{p_1}$ denotes the parity-check matrix of $\mathcal{C}_{p_1}$. Then 
\begin{align*}
    G H^T&=\left(\begin{bmatrix}
    1 & 1 & \cdots & 1
    \end{bmatrix}\otimes I_{p_1^{a_1-1}}\right)\left(H_{p_1}^T\otimes I_{p_1^{a_1-1}}\right)\\
    &=\begin{bmatrix}
    1 & 1 & \cdots & 1
    \end{bmatrix} H_{p_1}^T \otimes I_{p_1^{a_1-1}}I_{p_1^{a_1-1}}\\
    &=O\otimes I_{p_1^{a_1-1}}\\
    &=O.
\end{align*}
Since $\rank(H)=\dim \mathcal{C}_{p_1^{a_1}}^{\perp},$ the matrix $H$ is a generator matrix of $\mathcal{C}_{p_1^{a_1}}^{\perp}$. Thus, the code $\mathcal{C}_{p_1^{a_1}}^{\perp}$ is permutation equivalent to a code generated by the matrix
\begin{align*}
    I_{p_1^{a_1-1}}\otimes H_{p_1}=\begin{bmatrix}
        H_{p_1} & 0 & 0 & \cdots & 0\\
        0 & H_{p_1} & 0 & \cdots & 0\\
        \vdots & \vdots & \vdots & \ddots & \vdots\\
         0 & 0 & \cdots & 0 & H_{p_1}\\
    \end{bmatrix}.
\end{align*}
It is clear that the code generated by $I_{p_1^{a_1-1}}\otimes H_{p_1}$ is $\underbrace{\mathcal{C}_{p_1}^{\perp}\oplus\mathcal{C}_{p_1}^{\perp}\oplus\cdots \oplus \mathcal{C}_{p_1}^{\perp}}_{p_1^{a_1-1}\;\text{copies}}$, which has minimum distance $2$. Hence, the result follows.
\end{proof}
\begin{theorem}\label{thm:: main theorem}
    Let $n_1, n_2>1$ be integers co-prime to $\Char(\mathbb{F}_q).$ If $\gcd(n_1, n_2)=1,$ then $\mathcal{C}_{n_1 n_2}^{\perp}$ and $\mathcal{C}_{n_1}^{\perp}\otimes\mathcal{C}_{n_2}^{\perp}$ are permutation equivalent.
\end{theorem}
\begin{proof}
    To prove the theorem, it is enough to prove that $\mathcal{C}_{n_1 n_2}^{\perp}=\Psi(\mathcal{C}_{n_1}^{\perp}\otimes\mathcal{C}_{n_2}^{\perp}).$ Clearly, their lengths are the same. It is easy to see that the zeros of $Q_{n_1}(x)$ are the nonzeros of $\mathcal{C}_{n_1}^{\perp}$ and the zeros of $Q_{n_2}(x)$ are the nonzeros of $\mathcal{C}_{n_2}^{\perp}.$ Let $\alpha$ be a primitive $n_1$-th root of unity and $\beta$ be a primitive $n_2$-th root of unity. Then $\{ \alpha^{i}: 1\le i< n_1\; \text{and}\; \gcd(i, n_1)=1\}$ and $\{ \beta^{j}: 1\le j< n_2\; \text{and}\; \gcd(j, n_2)=1\}$ are the set of nonzeros of $\mathcal{C}_{n_1}^{\perp}$ and $\mathcal{C}_{n_2}^{\perp},$ respectively. Invoking Lemma \ref{Lem:: set of nonzeros}, $\mathcal{A}:=\{ \alpha^{i}\beta^{j}: 1\le i, j< n_1, \;\gcd(i, n_1)=1\; \text{and}\; \gcd(j, n_2)=1\}$ is the set of nonzeros of $\Psi(\mathcal{C}_{n_1}^{\perp}\otimes\mathcal{C}_{n_2}^{\perp}).$ 
    
    If $\alpha^i$ is a primitive $n_1$-th root of unity and $\beta^j$ is a primitive $n_2$-th root of unity, then $\alpha^i\beta^j$ is a primitive $n_1n_2$-th root of unity and hence, a root of $Q_{n_1n_2}(x).$ Consequently, $\mathcal{A}$ is the set of nonzeros of $\mathcal{C}_{n_1n_2}^{\perp}.$ Since $\dim \Psi(\mathcal{C}_{n_1}^{\perp}\otimes\mathcal{C}_{n_2}^{\perp})=\dim \mathcal{C}_{n_1}^{\perp} \dim \mathcal{C}_{n_2}^{\perp}=\varphi(n_1)\varphi(n_2)=\varphi(n_1n_2)=\dim \mathcal{C}_{n_1 n_2}^{\perp},$ they have exactly $\#\mathcal{A}$ nonzeros. Thus, we have $\mathcal{C}_{n_1 n_2}^{\perp}=\Psi(\mathcal{C}_{n_1}^{\perp}\otimes\mathcal{C}_{n_2}^{\perp}).$ 
\end{proof}

\begin{theorem}\label{thm:: for any n}
Let $n>1$ be a positive integer such that $\Char(\mathbb{F}_q) \nmid n.$ Then $\mathcal{C}_n^{\perp}$, the dual of $\mathcal{C}_n$, has parameters $[n,\varphi(n), 2^{\omega(n)}]_q$ over $\mathbb{F}_q.$
\end{theorem}

\begin{proof}
The length and the dimension follow directly from the definition of $\mathcal{C}_n$. 

If $n$ is a prime power, then the result follows from Theorem \ref{thm:: n is a single prime power}. Now let 
\[
n=p_1^{a_1}p_2^{a_2}\cdots p_k^{a_k}
\]
be the prime factorization of $n$, where $k\ge 2$. By Theorem \ref{thm:: n is a single prime power}, we have 
\[
d(\mathcal{C}_{p_i^{a_i}}^{\perp})=2,\; \textnormal{for all}\; 1\le i\le k. 
\]

Applying Theorem \ref{thm:: main theorem}, we obtain
\[
d(\mathcal{C}_{p_1^{a_1}p_2^{a_2}}^{\perp})
= d(\mathcal{C}_{p_1^{a_1}}^{\perp})\, d(\mathcal{C}_{p_2^{a_2}}^{\perp})
= 2 \times 2 = 4.
\]
Repeating this argument inductively and using Theorem \ref{thm:: main theorem}, we obtain
$d(\mathcal{C}_n^{\perp}) = 2^k = 2^{\omega(n)}$.
\end{proof}
We now turn our attention to the code $\mathcal{C}_{n,1}^{\perp},$ the Euclidean dual of $\mathcal{C}_{n,1}$. We begin with the following lemma.
\begin{lemma}\label{lemma: n-phi n}
    A composite number $n$ satisfies $n-\varphi(n)\ge 2^{\omega(n)}.$
\end{lemma}
\begin{proof}
    We prove by induction on $k=\omega(n)$.
    
    For $k=1$, $n=p_1^{a_1}$ for some prime $p_1$ and $a_1\ge 2$. Then $$n-\varphi(n)=p_1^{a_1-1}\ge 2=2^{\omega(n)}.$$
    
    For $k=2$, $n=p_1^{a_1}p_2^{a_2}$ for two distinct primes $p_1$ and $p_2$ and $a_1, a_2\ge 1$. Then $n-\varphi(n)=p_1^{a_1-1} p_2^{a_2-1}(p_1+p_2-1)\ge 2+3-1=4=2^{\omega(n)}$.
    
    Assume that the result holds for all composite $n$ with $\omega(n)=k\ge 2$.
    Let $N=n p^{a}$, where $p$ is a prime that does not divide $n$ and $a\ge 1$. Then $$\varphi(N)=\varphi(n)\varphi(p^{a})=\varphi(n)(p^{a}-p^{a-1})\le \varphi(n)p^{a}.$$
    Consequently, $N-\varphi(N)\ge n p^{a}-\varphi(n)p^{a}=(n-\varphi(n))p^{a}\ge 2^{k}\times 2= 2^{k+1}=2^{\omega(N)}.$
\end{proof}
\begin{lemma}\label{lemma: decomposition of Cn1}
    Let $n$ be a composite number such that $\Char(\mathbb{F}_q) \nmid n.$ Then $\mathcal{C}_{n,1}^{\perp}=\mathcal{C}_{n}^{\perp}+\mathcal{R}_n$.
\end{lemma}
\begin{proof}
    Let $h(x):=\frac{x^n-1}{Q_n(x)(x-1)}=\underset{\underset{d\notin \{1, n\}}{d\mid n}}{\prod} Q_d(x).$ Since $Q_d^*(x)=Q_d(x)$ for all $d\ge 2,$ we have $h^*(x)=h(x).$ Also note that, $h(0)=1.$ If $g^{\bot}(x)$ denotes the generator polynomial of $\mathcal{C}_{n,1}^{\bot},$ then $g^{\bot}(x):=h^{*}(x)/h(0)=h(x).$
    
    Since the sum of two cyclic codes of the same length is cyclic, the code $\mathcal{C}_{n}^{\perp}+\mathcal{R}_n$ is cyclic. Thus, it is enough to show that the generator polynomial of $\mathcal{C}_{n}^{\perp}+\mathcal{R}_n$ and $\mathcal{C}_{n,1}^{\perp}$ are the same. Since $\mathcal{C}_n^{\perp}=\left\langle \frac{x^n-1}{Q_n(x)} \right\rangle$ and $\mathcal{R}_n= \left\langle \frac{x^n-1}{x-1} \right\rangle$, $\mathcal{C}_{n}^{\perp}+\mathcal{R}_n=\left\langle \gcd\left(\frac{x^n-1}{Q_n(x)}, \frac{x^n-1}{x-1}\right) \right\rangle=\left\langle \frac{x^n-1}{Q_n(x) (x-1)} \right\rangle=\mathcal{C}_{n,1}^{\perp}.$
\end{proof}
\begin{remark}\label{Remark: main remark}
    Due to Theorem \ref{thm:: main theorem} and Lemma \ref{lemma: decomposition of Cn1}, the code $\mathcal{C}_{n,1}^{\perp}$ is permutation equivalent to the code $\left(\mathcal{C}_{p_1^{a_1}}^{\perp}\otimes \mathcal{C}_{p_2^{a_2}}^{\perp}\otimes\cdots \otimes \mathcal{C}_{p_k^{a_k}}^{\perp}\right) + \mathcal{R}_n$, where $n=p_1^{a_1}p_2^{a_2}\cdots p_k^{a_k}$ is the prime factorization of $n$.
\end{remark}
\begin{theorem}\label{thm: main thm 2}
    Let $n$ be a composite number such that $\Char(\mathbb{F}_q) \nmid n.$ Then $\mathcal{C}_{n,1}^{\perp},$ the dual of $\mathcal{C}_{n,1},$ has parameters $[n,\varphi(n)+1, 2^{\omega(n)}]_q$ over $\mathbb{F}_q.$
\end{theorem}
\begin{proof}
    The length and the dimension follow directly from the definition of $\mathcal{C}_{n,1}$. 
     Since $\mathcal{C}_{n,1}^{\perp}\supseteq\mathcal{C}_{n}^{\perp}$, $d(\mathcal{C}_{n,1}^{\perp})\le d(\mathcal{C}_{n}^{\perp})=2^{\omega(n)}$.
     
     Let $n=p_1^{a_1}p_2^{a_2}\cdots p_k^{a_k}$ be the prime factorization of $n$ and define
     $$\mathcal{D}_n:=\left(\mathcal{C}_{p_1^{a_1}}^{\perp}\otimes \mathcal{C}_{p_2^{a_2}}^{\perp}\otimes\cdots \otimes \mathcal{C}_{p_k^{a_k}}^{\perp}\right) + \mathcal{R}_n.$$
     Due to Remark \ref{Remark: main remark}, to show that $d(\mathcal{C}_{n,1}^{\perp})\ge 2^{\omega(n)}$, it is enough to prove that weight of every nonzero codeword of $\mathcal{D}_n$ is at least $2^{\omega(n)}$.

     Note that $$\mathcal{D}_n=\left(\mathcal{C}_{p_1^{a_1}}^{\perp}\otimes \mathcal{C}_{p_2^{a_2}}^{\perp}\otimes\cdots \otimes \mathcal{C}_{p_k^{a_k}}^{\perp}\right) \bigcup \left\{\bm{c}+\lambda \bm{1}: \bm{c}\in \mathcal{C}_{p_1^{a_1}}^{\perp}\otimes \mathcal{C}_{p_2^{a_2}}^{\perp}\otimes\cdots \otimes \mathcal{C}_{p_k^{a_k}}^{\perp}\; \text{and}\; \lambda\in \mathbb{F}_q^{\times}\right\}.$$ If a codeword of $\mathcal{D}_n$ belongs to $\mathcal{C}_{p_1^{a_1}}^{\perp}\otimes \mathcal{C}_{p_2^{a_2}}^{\perp}\otimes\cdots \otimes \mathcal{C}_{p_k^{a_k}}^{\perp}$, then its weight is at least $2^{\omega(n)}$, due to Theorem \ref{thm:: main theorem}.
     
     Let $\bm{c}+\lambda \bm{1}\in \mathcal{D}_n\setminus \{\bm{0}\}$, for some $\bm{c}\in \mathcal{C}_{p_1^{a_1}}^{\perp}\otimes \mathcal{C}_{p_2^{a_2}}^{\perp}\otimes\cdots \otimes \mathcal{C}_{p_k^{a_k}}^{\perp}$ and $\lambda\in \mathbb{F}_q^{\times}.$ The $i$-th coordinate $(\bm{c}+\lambda \bm{1})_i$ of $\bm{c}+\lambda \bm{1}$ is $0$ if and only if $c_i=-\lambda.$ So, for $\bm{c}\in\mathcal{C}_{p_1^{a_1}}^{\perp}\otimes \mathcal{C}_{p_2^{a_2}}^{\perp}\otimes\cdots \otimes \mathcal{C}_{p_k^{a_k}}^{\perp}$ and $\lambda\in \mathbb{F}_q^{\times}$, the minimum weight of $\bm{c}+\lambda \bm{1}$ is achieved by maximizing the number of coordinates $i$ for which $c_i=-\lambda$.

     For each $1\le i\le k$, $\mathcal{C}_{p_i}$ is the repetition code of length $p_i$ so that $\mathcal{C}_{p_i}^{\perp}:=\{\bm{x}=(x_1, x_2\dots, x_{p_i})\in \mathbb{F}_q^{p_i}: \sum_{j=1}^{p_i} x_j=0\}$. Since each $p_i$ is co-prime to $\Char(\mathbb{F}_q)$, the only codeword in $\mathcal{C}_{p_i}^{\perp}$ whose coordinates are all equal is the zero codeword. Hence, in any nonzero codeword of $\mathcal{C}_{p_i}^{\perp}$, at most $p_i-1$ coordinates can take the same value. This bound is attained, for example, by taking $x_j=\alpha\in\mathbb{F}_q$ for $1\le j\le p_i-1$ and $x_{p_i}=-(p_i-1)\alpha$. Since $\mathcal{C}_{p_i^{a_i}}^{\perp}$ is permutation equivalent to $\underbrace{\mathcal{C}_{p_i}^{\perp}\oplus\mathcal{C}_{p_i}^{\perp}\oplus\cdots \oplus \mathcal{C}_{p_i}^{\perp}}_{p_i^{a_i-1}\;\textnormal{copies}}$, the maximum number of coordinates that can take the same value in a nonzero codeword of $\mathcal{C}_{p_i^{a_i}}^{\perp}$ is $(p_i-1)p_i^{a_i-1}=\varphi(p_i^{a_i}).$ It follows that in the direct product $\mathcal{C}_{p_1^{a_1}}^{\perp}\otimes \mathcal{C}_{p_2^{a_2}}^{\perp}\otimes\cdots \otimes \mathcal{C}_{p_k^{a_k}}^{\perp}$, the maximum number of coordinates that can take the same value in a nonzero codeword is $\varphi(p_1^{a_1})\varphi(p_2^{a_2})\cdots \varphi(p_k^{a_k})=\varphi(n)$.

     Thus, $\min\left\{\wt(\bm{c}+\lambda \bm{1}): \bm{c}\in \mathcal{C}_{p_1^{a_1}}^{\perp}\otimes \mathcal{C}_{p_2^{a_2}}^{\perp}\otimes\cdots \otimes \mathcal{C}_{p_k^{a_k}}^{\perp}\; \text{and}\; \lambda\in \mathbb{F}_q^{\times}\right\}\ge n-\varphi(n)\ge 2^{\omega(n)}$, due to Lemma \ref{lemma: n-phi n}. As a result, $d(\mathcal{D}_n)\ge 2^{\omega(n)}$. Since $\mathcal{D}_n$ and $\mathcal{C}_{n,1}^{\perp}$ are permutation equivalent, the result follows.
\end{proof}
\section{Conclusion}\label{conclusion}
In this article, we investigated the Euclidean duals of the cyclic codes $\mathcal{C}_n$ and $\mathcal{C}_{n,1}$ generated by the $n$-th cyclotomic polynomial $Q_n(x)$ and the product $Q_n(x)Q_1(x)$, respectively over $\mathbb{F}_q$, where $n$ is a natural number co-prime to $q$. We provided a detailed description of the structure of these dual codes and, as a direct consequence, derived explicit formulas for their minimum distances as functions of $n$.

Moreover, our results resolve the conjectures posed in \cite[Conjecture 4.9 and Conjecture 4.10]{BHAGAT2025}, establishing that the minimum distance of both $\mathcal{C}_n^{\perp}$ and $\mathcal{C}_{n,1}^{\perp}$ is precisely $2^{\omega(n)}$. These findings not only confirm our earlier predictions but also offer a clear structural understanding of the duals of these codes.

\section*{Declarations}
\subsection*{Conflict of Interest} Both authors declare that they have no conflict of interest.

\subsection*{Acknowledgements}
The authors are grateful to Professor Markus Grassl for providing valuable insight that helped guide the direction of this work.

The work of the first author was supported by the Council of Scientific and Industrial Research (CSIR) India, under grant no. 09/0086(13310)/2022-EMR-I.

\bibliographystyle{abbrv}
\bibliography{Cyclo_bibliography}

\end{document}